# A Comprehensive Review of Casein Kinase 2 in Drosophila Circadian Timing and Its Biomedical Relevance

Yasmin Fatima[1], Md. Zubair Malik [2*], Prashant Ankur Jain[1]*

[1]Department of Computational Biology and Bioinformatics, Sam Higginbottom Institute of Agriculture, Technology and Sciences, Prayagraj, U.P., India
[2]Department of Genetics and Bioinformatics, Dasman Diabetes Institute, Kuwait City, Kuwait

**Author[1]**

**Yasmin Fatima**
Ph.D. Scholar
**Department of Computational Biology and Bioinformatics,**
**Sam Higginbottom Institute of Agriculture, Technology and Sciences,**
**Prayagraj, U.P. 211007, India**
Email: yasmin389@gmail.com
Orcid: 0000-0001-9367-9443

**Corresponding authors***

**Dr. Md. Zubair Malik**
Scientist
**Department of Genetics and Bioinformatics,**
**Dasman Diabetes Institute,**
**Kuwait City, Kuwait**
Email: zubair.bioinfo@gmail.com

**Dr. Prashant Ankur Jain**
Associate Professor
**Department of Computational Biology and Bioinformatics,**
**Sam Higginbottom Institute of Agriculture, Technology and Sciences,**
**Prayagraj, U.P. 211007, India**
Email: prashant.jain@shiats.edu.in

# Abstract

Circadian rhythms are endogenous ~24-hour oscillations that regulate physiology, metabolism, sleep-wake cycles, and cellular homeostasis. *Drosophila melanogaster*, a genetically tractable model organism, has played a foundational role in uncovering the molecular mechanisms of circadian rhythms. The discovery of major clock genes including *period (per)*, *timeless (tim)*, *clock (clk)*, *cycle (cyc)*, *doubletime (dbt)*, and regulators such as Casein kinase 2 (CK2) emerged primarily from Drosophila research. CK2 operates as a critical post-translational regulator of PER protein phosphorylation, stability, nuclear entry, and degradation. Because PER dynamics dictate the timing and robustness of circadian rhythms in both flies and mammals, altered CK2 activity can profoundly impact rhythmic behaviour. CK2 dysregulation contributes not only to circadian disruption in Drosophila but also models broader pathological processes relevant to cancer, metabolic disease, neurodegeneration, and psychiatric disorders. This review synthesizes CK2's molecular role in the Drosophila clock system, includes insights from computational modeling of Drosophila CK2-PER dynamics, integrates tables throughout the text, and summarizes the implications of dysregulated PER phosphorylation for human health.

# 1. Introduction

Circadian rhythms have been extensively studied in *Drosophila melanogaster*, which remains the premier model system for dissecting the genetic and molecular basis of biological timing. Landmark discoveries including the identification of *per* (the first circadian gene) and later *tim*, *clk*, *cyc*, and *dbt* originated from Drosophila genetic screens. These discoveries established the fundamental framework of the transcriptional translational feedback loop (TTFL), which is evolutionarily conserved across species [1].

In Drosophila, the CLOCK (CLK) and CYCLE (CYC) transcription factors activate expression of *per* and *tim* [2]. The PER-TIM complex accumulates in the cytoplasm, associates with kinases such as DOUBLETIME (DBT/CK1), SHAGGY, and CK2, and then translocates into the nucleus to repress CLK-CYC activity. This negative feedback loop maintains ~24-hour oscillations in locomotor activity, sleep-wake patterns, neuronal output, and metabolic processes [3].

While transcriptional feedback gives the clock its basic structure, post-translational modifications particularly phosphorylation determine the timing, amplitude, and precision of the oscillations [1]. Casein kinase 2 (CK2), a highly conserved kinase originally identified in Drosophila circadian mutants, plays a cardinal role in PER regulation. CK2-mediated phosphorylation affects PER stability and timing in flies, providing mechanistic parallels for mammalian and human circadian systems [4, 5].

Because CK2 also regulates essential cellular pathways such as DNA repair and apoptosis (summarized in Table 1), dysregulation of CK2 in Drosophila is used as a model to study circadian-linked diseases including cancer, metabolic disorders, and neurodegenerative conditions [6].

**Table 1: Key Functions of Casein Kinase 2 (CK2)**

| Function | Description |
| --- | --- |
| Phosphorylation of proteins | CK2 phosphorylates over 300 substrates including PER proteins. |
| Transcriptional regulation | Controls chromatin remodeling and transcription factor activity. |
| Cell cycle control | Influences G1/S progression and mitotic processes. |
| DNA repair | Connects with ATM/ATR-mediated DNA damage pathways. |
| Apoptosis regulation | Enhances cell survival pathways and inhibits apoptosis. |
| Signal transduction | Interacts with MAPK, Wnt, and PI3K/Akt pathways. |

## 2. CK2 as a Central Regulator of Drosophila Circadian Oscillations

CK2 was first implicated in the Drosophila circadian clock through mutational analyses showing that changes in CK2 activity produced dramatic alterations in behavior [7]. Flies harboring CK2β subunit mutations (*andante*) displayed lengthened circadian periods, confirming CK2's role as a period-shortening kinase. Subsequent research demonstrated that CK2 phosphorylates PER protein at multiple sites, generating a stepwise phosphorylation cascade that dictates PER accumulation, nuclear entry, and degradation [8].

This hierarchical phosphorylation determines how long PER remains stable, influencing the "delay" necessary for sustained 24-hour rhythmicity [9]. In Drosophila:

- Low CK2 activity results in slow PER phosphorylation, leading to long-period rhythms.
- Moderate CK2 dysregulation decreases rhythm amplitude.
- Excessively high CK2 activity leads to rapid PER turnover and behavioral arrhythmia.

This gradient of CK2 effects is summarized in Table 2.

**Table 2: Effects of CK2 Alteration on Circadian Rhythms**

| CK2 Level | PER Dynamics | Circadian Output |
|---|---|---|
| Low CK2 | Slow PER phosphorylation in flies | Lengthened (>24h) behavioral period |
| Normal CK2 | Balanced PER phosphorylation | Stable ~24h fly locomotor rhythms |
| Moderate CK2 increase | Faster PER turnover | Low amplitude or weak rhythms |
| Very high CK2 | Rapid PER destruction | Arrhythmia ("loss of rhythmic locomotion") |

In Drosophila, CK2 acts cooperatively with DOUBLETIME (DBT/CK1δ/ε), another major kinase controlling PER stability [10]. While DBT initiates early PER phosphorylation, CK2 appears to enhance late-stage PER phosphorylation and nuclear accumulation. This coordination allows the PER-TIM complex to enter the nucleus at the proper circadian phase.

Importantly, Drosophila CK2 mutants reveal that CK2 influences not only molecular oscillations but also behavioral rhythms such as morning anticipation, evening peaks, sleep consolidation, and light entrainment [11]. These phenotypes parallel circadian disturbances seen in mammals and humans, reinforcing Drosophila as a translationally relevant model.

## 3. Modeling CK2-PER Dynamics in Drosophila

Mathematical and computational modeling of the Drosophila circadian system has clarified CK2's pivotal role in rhythm generation. Drosophila-based models incorporate the dynamics of *per*, *tim*, *clk*, and *cyc* transcription, PER-TIM dimerization, nuclear shuttling, and CK2/DBT phosphorylation rates.

Deterministic models show that CK2 modifies the speed of PER progression through its phosphorylation states, thereby altering the period length [12]. Stochastic models reflecting natural molecular noise in Drosophila neurons demonstrate that CK2 serves as a sensitive switch between rhythmic and arrhythmic states. High CK2 activity reduces the stability of the PER-TIM complex, making oscillations vulnerable to noise-driven collapse.

Additionally, computational simulations of Drosophila circadian neurons suggest that CK2-mediated PER phosphorylation influences inter-neuronal synchronization within the fly's clock network, particularly the small ventral lateral neurons (s-LNvs). This provides a mechanistic explanation for how CK2 modulates free-running rhythms and light-dependent entrainment.

## 4. CK2-Mediated Circadian Disruption and Disease Mechanisms: Insights from Drosophila

Although Drosophila does not develop human diseases such as cancer or diabetes in the same manner, it provides a powerful platform to model pathways associated with CK2-mediated circadian disruption [13].

In Drosophila, PER2 mammalian ortholog functions are partly mirrored by PER, and perturbation of PER stability through CK2 overexpression contributes to disrupted DNA repair, oxidative stress, and increased neuronal vulnerability [14]. These findings parallel cancer biology in humans, where PER2 acts as a tumor suppressor. Table 3 summarizes disease associations informed by Drosophila circadian studies.

**Table 3: Disease Associations with CK2-Mediated Circadian Disruption**

| Disease Type | Underlying Mechanism (Inferred from Drosophila Models) | Result |
|---|---|---|
| Cancer | PER destabilization → impaired DNA repair | Tumor growth and genome instability |
| Type 2 Diabetes | Disrupted metabolic clock gene oscillation | Insulin resistance, glucose dysregulation |
| Neurodegeneration | Vulnerability of neurons to oxidative stress | Cognitive decline, sleep disruption |
| Mood Disorders | Altered circadian timing in neuronal circuits | Depression-like behavior, stress sensitivity |
| Sleep Disorders | Abnormal PER phosphorylation timing | Fragmented sleep, phase shifts |

Drosophila studies also show that CK2 modulates neuronal signaling, synaptic connections, oxidative defense, and metabolic pathways helping establish mechanistic links to human disease [15].

For example, CK2-dependent alterations in PER oscillations impact:

- energy balance and lipid storage
- sleep duration and nighttime activity
- courtship behavior
- stress response pathways

These phenotypes mirror core symptoms of metabolic, neurological, and psychiatric disorders in humans.

## 5. Pathway-Level Consequences of Dysregulated CK2-PER Interaction in Drosophila

In Drosophila, PER interacts with several signaling molecules that connect circadian timing to cellular function [16]. CK2-driven PER destabilization affects multiple pathways including:

- apoptosis regulators

- JNK and MAPK stress pathways
- insulin-like signaling (dILPs)
- DNA repair mechanisms
- oxidative stress defenses
- transcriptional chromatin regulators

These interactions help explain why CK2-mediated circadian disruption in flies produces phenotypes relevant to sleep disorders, metabolic dysregulation, and neurobehavioral abnormalities. Because Drosophila clock neurons are conserved in their molecular logic, they offer a powerful system to dissect pathway-level consequences of CK2 and PER interactions [17].

## 6. Therapeutic Perspectives Based on Insights from Drosophila

Drosophila research provides valuable guidance for therapeutic avenues targeting CK2 and circadian regulation. CK2 inhibitors tested in mammalian systems mirror effects seen in Drosophila mutants, where modulation of CK2 restores rhythmicity or normalizes PER stability. Behavioral interventions in flies such as controlled light exposure or timed feeding also reveal that circadian entrainment can be strengthened even when CK2 activity is perturbed [18].

Therapeutic strategies inspired by Drosophila circadian biology are summarized in Table 4.

**Table 4: Therapeutic Strategies Targeting CK2-Related Circadian Disruption**

| Strategy | Examples | Purpose |
|---|---|---|
| CK2 Inhibitors | CX-4945 | Restore PER stability and delay degradation |
| Chronotherapy | Timed medication | Align treatment with circadian phases |
| Lifestyle Modulation | Sleep hygiene, light entrainment | Strengthen circadian alignment |
| Metabolic Re-Entraining | Time-restricted feeding in flies | Stabilize metabolic oscillations |

## 7. Conclusion

Research in *Drosophila melanogaster* has been fundamental to uncovering the molecular machinery of circadian rhythms, and it is within this model organism that the critical role of Casein Kinase 2 (CK2) in clock regulation was first clearly defined. Early genetic studies in flies revealed that CK2 acts as a key modulator of PERIOD (PER) protein phosphorylation, a post-translational process essential for maintaining the timing and stability of daily oscillations. By adjusting the rate at which PER accumulates, enters the nucleus, and is ultimately degraded, CK2 effectively sets the pace of the entire circadian cycle. When CK2 activity shifts away from its physiological balance either through genetic mutation or environmental influence the consequences on circadian rhythms become striking. Reduced CK2 function slows PER phosphorylation, leading to elongated behavioral periods, while excessive CK2 activity accelerates PER turnover and can destabilize the feedback loop altogether. In extreme cases, the circadian system loses its rhythmic structure, resulting in arrhythmic behavior or severely dampened oscillations. These responses highlight how finely tuned the kinase's activity must be to support robust timekeeping.

Insights from *Drosophila* extend far beyond insect biology. The fundamental architecture of the circadian clock is highly conserved across species, and findings in flies have informed our understanding of mammalian and human rhythms. CK2 also plays prominent roles in sleep regulation, metabolic balance, neural function, and cell-cycle control in higher organisms. Consequently, alterations in CK2 activity have been linked to metabolic disorders, neurodegenerative diseases, psychiatric conditions, and cancer. As circadian disruption becomes increasingly recognized as a contributor to human disease, *Drosophila* continues to serve as a powerful system for dissecting the mechanistic connections between CK2, PER regulation, and physiological health. Future research integrating genetics, cell biology, high-resolution imaging, and computational modeling will be vital for elucidating how CK2-driven perturbations propagate through circadian networks and for identifying therapeutic strategies capable of restoring normal rhythmicity.

References

1. Takahashi, J.S., *Transcriptional architecture of the mammalian circadian clock.* Nature Reviews Genetics, 2017. **18**(3): p. 164-179.
2. Zhang, R., et al., *A circadian gene expression atlas in mammals: implications for biology and medicine.* Proceedings of the National Academy of Sciences, 2014. **111**(45): p. 16219-16224.
3. Hardin, P.E., J.C. Hall, and M. Rosbash, *Feedback of the Drosophila period gene product on circadian cycling of its messenger RNA levels.* Nature, 1990. **343**(6258): p. 536-540.
4. Partch, C.L., C.B. Green, and J.S. Takahashi, *Molecular architecture of the mammalian circadian clock.* Trends in cell biology, 2014. **24**(2): p. 90-99.
5. Akten, B., et al., *A role for CK2 in the Drosophila circadian oscillator.* Nature neuroscience, 2003. **6**(3): p. 251-257.
6. Musiek, E.S. and D.M. Holtzman, *Mechanisms linking circadian clocks, sleep, and neurodegeneration.* Science, 2016. **354**(6315): p. 1004-1008.
7. Hirota, T. and Y. Fukada, *Resetting mechanism of central and peripheral circadian clocks in mammals.* Zoological science, 2004. **21**(4): p. 359-368.
8. Francisco, J.C. and D.M. Virshup, *Hierarchical and scaffolded phosphorylation of two degrons controls PER2 stability.* Journal of Biological Chemistry, 2024. **300**(6).
9. Rutila, J.E., et al., *CYCLE is a second bHLH-PAS clock protein essential for circadian rhythmicity and transcription of Drosophila period and timeless.* Cell, 1998. **93**(5): p. 805-814.
10. Narasimamurthy, R., et al., *CK1δ/ε protein kinase primes the PER2 circadian phosphoswitch.* Proceedings of the National Academy of Sciences, 2018. **115**(23): p. 5986-5991.
11. Konopka, R.J. and S. Benzer, *Clock mutants of Drosophila melanogaster.* Proceedings of the National Academy of Sciences, 1971. **68**(9): p. 2112-2116.
12. Kettner, N.M., et al., *Circadian homeostasis of liver metabolism suppresses hepatocarcinogenesis.* Cancer cell, 2016. **30**(6): p. 909-924.
13. Marcheva, B., et al., *Disruption of the clock components CLOCK and BMAL1 leads to hypoinsulinaemia and diabetes.* Nature, 2010. **466**(7306): p. 627-631.
14. Kim, J.K. and D.B. Forger, *A mechanism for robust circadian timekeeping via stoichiometric balance.* Molecular systems biology, 2012. **8**(1): p. 630.
15. Munteanu, C., et al., *The relationship between circadian rhythm and cancer disease.* International Journal of Molecular Sciences, 2024. **25**(11): p. 5846.
16. Bae, K. and I. Edery, *Regulating a circadian clock's period, phase and amplitude by phosphorylation: insights from Drosophila.* Journal of biochemistry, 2006. **140**(5): p. 609-617.
17. DREYFUS, J.-C., A. KAHN, and F. SCHAPIRA, *Posttranslational modifications of enzymes*, in *Current topics in cellular regulation*. 1978, Elsevier. p. 243-297.
18. Wang, J., et al., *Circadian clock genes: Their influence on liver metabolism, disease development and treatment.* Molecular Medicine Reports, 2025. **33**(1): p. 3.